\documentclass[aps, prd, twocolumn, groupedaddress]{revtex4-2}

\makeatletter

\usepackage{amsmath,amssymb,amsfonts,bm,braket,cases,mathrsfs}
\usepackage{graphicx}
\renewcommand{\vec}[1]{\bm{#1}} 
\newcommand{\mat}[1]{\mathbf{#1}}   
\newcommand{\flavor}[1]{\mathsf{#1}}    

\newcommand{\re}{\operatorname{Re}}
\newcommand{\im}{\operatorname{Im}}
\newcommand{\diag}{\operatorname{diag}}
\newcommand{\adj}{\operatorname{adj}}

\makeatother

\begin{document}

\preprint{APS/123-QED}

\title{Spatiotemporal linear instability analysis of collective neutrino flavor conversion in 4-dimensional spacetime}

\author{Taiki Morinaga}

\affiliation{Graduate School of Advanced Science and Engineering, Waseda University, 3-4-1 Okubo, Shinjuku, Tokyo 169-8555, Japan.}

\date{\today}
\begin{abstract}
In an environment with high-density neutrinos formed in a core-collapse supernova (CCSN), the neutrinos exhibit nonlinear and complex oscillation behaviors due to their self-interactions.
The onset of this nonlinear oscillation can be investigated by linearizing the evolution equation for small perturbations around the flavor eigenstates.
While the condition under which the flavor eigenstates are unstable has been investigated in many studies, how the perturbations evolve in spacetime has yet to be elucidated.
In this paper, we analytically and correctly derive the asymptotic behaviors of the linear perturbations in 4-dimensional spacetime in the linear regime for a 2-beam neutrino model using the recently proposed Lefschetz thimble formulation.
The result suggests that the perturbations grow in the directions between the two neutrino beams.
We also briefly discuss the possible effects of neutrino flavor conversion on the explosion mechanism of a CCSN.
In particular, the result implies that the flavor instability in the preshock region may propagate into the 
postshock region, contrary to the previous study focusing on the group velocity in 1-dimensional space.
How to treat the case of a more realistic continuous spectrum is also discussed.
\end{abstract}
\maketitle

\section{Introduction}
Neutrino oscillation is a well-known phenomenon in which the survival probability of each neutrino flavor oscillates due to the deviations between the flavor eigenstates and mass eigenstates of neutrinos.
When background matter is considered, forward scattering on charged leptons changes the effective mass of the neutrinos and, hence, their oscillation behaviors.
This is called the Mikheyev-Smirnov-Wolfenstein (MSW) effect~\cite{Wolfenstein1978,Wolfenstein1979}, and its behavior is well understood.
Additionally, neutrino oscillation is similarly affected by neutrino self-interactions, which, unlike the MSW effect, transform neutrino oscillation into a nonlinear phenomenon.
This phenomenon is called collective neutrino oscillation, and its complex behaviors have been intensively studied by many researchers.

Collective neutrino oscillations are important only in environments with large amounts of neutrinos, such as supernovae and the early universe.
In particular, flavor conversion may affect the explosion mechanism of a core-collapse supernova (CCSN)~\cite{Takiwaki2014,Lentz2015,Janka2016,Oconnor2018,Ott2018,Vartanyan2018,Muller2018,Burrows2019}.
It is widely believed that the stagnant shock in a supernova core needs to be revived through the introduction of additional energy for the explosion to be successful.
Nucleons in the postshock region can obtain energy by absorbing $\nu_e$ and $\bar\nu_e$ emitted from neutrinospheres, while the absorption of the other flavors of neutrinos is almost kinematically prohibited.
Therefore, flavor conversion can change the effective heating rate of the shock and may play a key role in determining the success of the explosion.

Although numerical approaches to solving the nonlinear equations for collective neutrino oscillation suffer from enormous computational costs, attempts have been made to address these equations under certain assumptions in many studies.
For example, the time evolution of homogeneous monochromatic single-angle neutrinos can be understood as analogous to the motion of a pendulum and a spinning top and exhibits several interesting phenomena, such as synchronized and bipolar oscillations \cite{Hannestad2006,Johns2018}.
Numerical calculations treating many modes have also been performed and have revealed some new phenomena, such as spectral swaps or splitting \cite{Sawyer2005,Duan2006,Raffelt2007,Raffelt2007a,Esteban-Pretel2007,Esteban-Pretel2008,Dasgupta2008,Sawyer2009,Dasgupta2009a,Raffelt2010a,Duan2010,Yang2017,Dasgupta2018,Abbar2018b,Martin2019,Rrapaj2019,Abbar2020b,Shalgar2020,Bhattacharyya2020,Johns2020,Capozzi2020,Martin2021}.
It should be stressed that in all of these studies, the nonlinear differential equations have been solved by imposing certain symmetries to reduce their dimensionality in spacetime.

Linear stability analysis has also been applied to investigate the conditions for the onset of flavor conversion \cite{Banerjee2011,Mirizzi2012,Mirizzi2012a,Raffelt2013,Mirizzi2013,Chakraborty2014,Chakraborty2014a,Chakraborty2016,Sawyer2016,Dasgupta2017a,Izaguirre2017,Abbar2018,Abbar2018a,Airen2018,Yi2019,DelfanAzari2020,Glas2020,Morinaga2020a,Chakraborty2020}.
Such analysis involves considering small perturbations around the flavor eigenstates and searching for growing modes from the derived dispersion relations (DRs).
Flavor conversion can occur if there is a mode with a real wave vector $\vec{k}$ and a complex angular frequency $\omega$.
Many studies have predicted that flavor conversion occurs when the angular distribution of the electron lepton number (ELN) crosses 0, although this has not been proven mathematically.
Moreover, such linear analysis suggests the importance of multidimensionality.
Generally, the flavor eigenstates are more unstable in higher dimensions, and the results of flavor oscillation could drastically change in reduced dimensions.
Therefore, it is important to consider 4-dimensional spacetime to understand realistic flavor conversion behaviors.

Normal mode analysis can reveal the conditions for the onset of flavor conversion and which plane wave modes grow or decay.
In reality, however, perturbations may take the form of wave packets, and how they evolve in spacetime cannot be trivially determined from the DRs.
Recently, it has been suggested that the spatiotemporal behaviors of perturbations can be investigated via the method originally proposed by Briggs~\cite{Briggs1964,Landau1997,Capozzi2017}.
One can derive how perturbations grow from the coalescence features of the analytic continuation of the DR $k(\omega)$.
Instabilities can be classified as either absolute or convective instabilities on the basis of the behaviors of perturbations in the form of wave packets.

This analysis, however, assumes that a perturbation is a wave packet in only one direction and homogeneous in the other directions, which is not realistic in supernovae.
Recently, we developed a general and powerful method to investigate the spatiotemporal evolution of linear perturbations by using the Lefschetz thimble formulation \cite{Morinaga2020}.
This formulation can be used to treat multidimensional perturbations for arbitrary DRs.
In this paper, we apply this method to the collective neutrino flavor conversion in a 2-beam neutrino model and reveal the flavor conversion behaviors in 4-dimensional spacetime.
This is the first study to treat the spatiotemporal evolution of collective neutrino oscillations in 4-dimensional spacetime, although this is done in the linear regime.

The results for the 2-beam model show an absence of absolute instabilities in 4-dimensional spacetime, although they are present when only 2-dimensional perturbations are considered.
Moreover, the results suggest that flavor instabilities grow toward the directions between two neutrino beams and may imply the impact of flavor conversion on shock heating mechanisms in CCSNe.

This paper is organized as follows.
In Sect.~II, we linearize the kinetic equations that describe neutrino oscillations to treat perturbations around the flavor eigenstates.
In addition, we introduce the 2-beam neutrino model and reduce the linearized equations.
In Sect.~III, we perform a spatiotemporal analysis of the obtained linearized equations for the 2-beam model and discuss general cases other than the 2-beam model.
Sect.~IV concludes the paper.

\section{Kinetic equations for neutrinos}

\subsection{Kinetic equations for neutrinos}
We begin with the kinetic equation for the neutrino density matrix $\flavor{f}$ \cite{Sigl1993,Yamada2000,Cardall2008,Vlasenko,Izaguirre2017}:
\begin{align}
    v\cdot\partial\flavor{f}(x,\Gamma) = -i[\flavor{H}(x,\Gamma),\flavor{f}(x,\Gamma)] + \mathscr{C}[\flavor{f}],
    \label{eq:Kinetic}
\end{align}
which describes the evolution of streaming neutrinos in a potential generated by matter and neutrinos.
$(x^\mu)=(t,\vec{x})$ represents the coordinates in spacetime, and $\Gamma=(E,\vec{v})$ denotes the energy and flight direction of the neutrinos. We express antineutrinos by means of a density matrix with negative energy, as follows: $\flavor{f}(-E) \equiv -\bar{\flavor{f}}(E)\ (E > 0)$.
The Hamiltonian $\flavor{H}$ is given by
\begin{align}
    \flavor{H}(x,\Gamma) = \flavor{H}_{\mathrm{vac}}(E)+\flavor{H}_{\mathrm{int}}(x,\vec{v}),
\end{align}
where $\flavor{H}_{\mathrm{vac}}$ corresponds to vacuum oscillation and $\flavor{H}_{\mathrm{int}}$ to the potential.
The vacuum oscillation term is
\begin{align}
    \flavor{H}_{\mathrm{vac}}(E) \equiv \dfrac{\flavor{M}^{2}}{2E},
\end{align}
where $\flavor{M}^2$ is the neutrino mass-squared matrix.
The potential term is written as
\begin{align}
    \flavor{H}_{\mathrm{int}}(x,\vec{v}) \equiv v\cdot\flavor{\Lambda}(x)
\end{align}
with
\begin{align}
    \flavor{\Lambda}^\mu(x)\equiv \sqrt{2}G_F\left[\diag\left(\{j^\mu_\alpha(x)\}_\alpha\right) + \int d\Gamma \flavor{f}(x,\Gamma)v^\mu\right],
\end{align}
where $j_{\alpha}^\mu$ is a lepton number 4-current of charged lepton $\alpha$ and
\begin{align}
    \int d\Gamma \equiv \int_{-\infty}^\infty \dfrac{dE E^2}{2\pi^2}\int\dfrac{d^2\vec{v}}{4\pi}.
\end{align}
The collision term $\mathscr{C}[\flavor{f}]$ thermalizes the density matrix $\flavor{f}$, or causes the decoherence of neutrino oscillations in general.
The effects of this term on collective neutrino oscillations have recently been discussed \cite{Capozzi2018a,Shalgar,Martin2021}.
In this study, we neglect these effects because they are usually much smaller than the flavor instability induced by neutrino potentials in supernovae.

\subsection{Linearization}
Here, we consider 2-flavor neutrinos $\nu_e$ and $\nu_x$, because each pair of flavors can be decoupled from the others even if we consider the number of flavors to be more than 2~\cite{Chakraborty2020}.
Then, the mass-squared matrix can be expressed as
\begin{align}
    \flavor{M}^{2} \equiv \flavor{U}
    \begin{pmatrix}
        m_{1}^{2} & 0\\
       0 & m_{2}^{2}
    \end{pmatrix}
    \flavor{U}^{-1},
\end{align}
with the neutrino mixing matrix
\begin{align}
    \flavor{U}\equiv
    \begin{pmatrix}
        \cos\theta & -\sin\theta\\
        \sin\theta & \cos\theta
    \end{pmatrix},
\end{align}
where $\theta$ is the mixing angle.
Additionally, the density matrix $\flavor{f}$ is a $2\times 2$ Hermitian matrix and can be decomposed by means of $(\flavor{\tau}_a)\equiv (\flavor{I}_2/2,\vec{\sigma}/2)$ with the Pauli matrices $\vec{\sigma}$ as follows:
\begin{align}
    \flavor{f}(x,\Gamma) = f^a(x,\Gamma)\flavor{\tau}_a.
\end{align}
The commutator of the Hermitian matrices $\flavor{A} = A^a \tau_a$ and $\flavor{B} = B^a \tau_a$ can be expressed as
\begin{align}
    [\flavor{A},\flavor{B}] =& A^a B^b[\flavor{\tau}_a,\flavor{\tau}_b]\\
    =& i\epsilon_{ab}^{\ \ c}A^a B^b \flavor{\tau}_c,
\end{align}
where $\epsilon_{ab}^{\ \ c}$ is the Levi-Civita symbol if $a$, $b$ and $c$ are permutations of $(1\ 2\ 3)$ and 0 otherwise.
Therefore, Eq.~(\ref{eq:Kinetic}) can be recast as
\begin{align}
    v\cdot\partial f^a(x,\Gamma) = \epsilon_{bc}^{\ \ a}H^b(x,\Gamma)f^c(x,\Gamma),
    \label{eq:KineticBloch}
\end{align}
where the decomposed Hamiltonian components are
\begin{align}
    H_{\mathrm{vac}}(E) = \dfrac{1}{2E}\begin{pmatrix}
        m_1^2+m_2^2 \\
        \Delta m^2\sin 2\theta\\
        0 \\
        \Delta m^2\cos 2\theta
    \end{pmatrix}
\end{align}
and
\begin{align}
    &H_{\mathrm{int}}(x,\vec{v}) \nonumber\\
    &= \sqrt{2}G_F v_\mu\left[\begin{pmatrix}
        j_e^\mu(x) + j_x^\mu(x) \\
        0\\
        0 \\
        j_e^\mu(x) - j_x^\mu(x)
    \end{pmatrix} + \int d\Gamma'f(x,\Gamma')v'{}^\mu
    \right],
\end{align}
with $\Delta m^2 \equiv m_1^2 - m_2^2$.

To investigate the onset of flavor conversion, we focus on the deviation from a flavor eigenstate and express the density matrix as
\begin{align}
    f(x,\Gamma)= \begin{pmatrix}
        f_0(x, \Gamma)\\
        0 \\
        0 \\
        f_\mathrm{c}(x, \Gamma)
    \end{pmatrix} + f_\mathrm{c}(x, \Gamma)\varepsilon(x,\Gamma),
\end{align}
where $f_{0}=f_{ee}+f_{xx}$ is the incoherent part, which does not play a role in flavor mixing, and $f_{\mathrm{c}}=f_{ee}-f_{xx}$ is the coherent part.
Substitution of this expression into Eq.~(\ref{eq:KineticBloch}) yields
\begin{align}
    v\cdot\partial\varepsilon(x,\Gamma) =
    &s(x, \Gamma)\nonumber\\
    &- c(x, \Gamma)\varepsilon(x, \Gamma)\nonumber\\
    &- \omega_\mathrm{s}(E)T_{23}\varepsilon(x,\Gamma)\nonumber\\
    &- \{\omega_\mathrm{c}(E) + v\cdot\Lambda_c(x)\}T_{12}\varepsilon(x,\Gamma)\nonumber\\
    &+ \sqrt{2}G_F \int d\Gamma'f_\mathrm{c}(\Gamma')v\cdot v'T_{12}\varepsilon(x,\Gamma')\nonumber\\
    &+ O(|\varepsilon|^2),
    \label{eq:KineticForPerturbation}
\end{align}
where
\begin{align}
    \omega_\mathrm{s}(E) \equiv \dfrac{\Delta m^2}{2E}\sin 2\theta,\ \omega_\mathrm{c}(E) \equiv \dfrac{\Delta m^2}{2E}\cos 2\theta,
\end{align}
\begin{align}
    \Lambda_\mathrm{c}^\mu(x) \equiv \sqrt{2}G_F\left[j_e^\mu(x)-j_x^\mu(x) + \int d\Gamma f_\mathrm{c}(\Gamma)v^\mu\right],
\end{align}
\begin{align}
    c(x, \Gamma) \equiv v\cdot\partial\ln f_\mathrm{c}(x, \Gamma)
\end{align}
and
\begin{align}
    T_{12} \equiv 
    \begin{pmatrix}
        0 & 0 & 0 & 0\\
        0 & 0 & 1 & 0\\
        0 & -1 & 0 & 0\\
        0 & 0 & 0 & 0
    \end{pmatrix},\ 
    T_{23} \equiv 
    \begin{pmatrix}
        0 & 0 & 0 & 0\\
        0 & 0 & 0 & 0\\
        0 & 0 & 0 & 1\\
        0 & 0 & -1 & 0
    \end{pmatrix}.
\end{align}
$s$ is the term of zeroth order in $\varepsilon$, expressed as
\begin{align}
    s(x, \Gamma) \equiv 
    \begin{pmatrix}    
        -\frac{v\cdot\partial f_0(x, \Gamma)}{f_\mathrm{c}(x, \Gamma)} \\
        0\\
        -\omega_\mathrm{s}(E)\\
        - c(x, \Gamma)\\
    \end{pmatrix}.
    \label{eq:seed}
\end{align}
This term appears because the flavor eigenstate is not a fixed point of Eq.~(\ref{eq:Kinetic}).
In the linear analysis we consider, we address the time evolution of $\varepsilon$ from the flavor eigenstate $\varepsilon(0,\vec{x}, \Gamma) = 0$.
Additionally, the spatial domain can be taken to be the open space $\mathbb{R}^3$, and the boundary condition for $\varepsilon$ is given by Eq.~(\ref{eq:KineticForPerturbation}) with $f_\mathrm{c} = 0$ and $\Lambda_\mathrm{c} = 0$ at $|\vec{x}|\to\infty$.
Under these initial and boundary conditions, the absence of $s$ yields $\varepsilon(x, \Gamma) = 0$ at all times; hence, $s$ is regarded as the seed perturbation.

Here, we define
\begin{align}
    S(x,\Gamma) \equiv \varepsilon^1(x,\Gamma) - i\varepsilon^2(x,\Gamma)
\end{align}
to recast Eq.~(\ref{eq:KineticForPerturbation}) as
\begin{widetext}
\begin{numcases}{}
v\cdot \partial\varepsilon^0(x,\Gamma) = -\frac{v\cdot\partial f_0(x, \Gamma)}{f_\mathrm{c}(x, \Gamma)} - c(x, \Gamma),
        \label{eq:LinearizedKinetic0}\\
\left[v\cdot \left\{i\partial -\Lambda_\mathrm{c}(x)\right\} - \omega_\mathrm{c}(E) + ic(x, \Gamma)\right]S(x,\Gamma) + \sqrt{2}G_F\int d\Gamma' f_\mathrm{c}(x, \Gamma')v\cdot v'S(x,\Gamma') + \omega_\mathrm{s}(E) \varepsilon^3(x, \Gamma) = -\omega_\mathrm{s}(E),
        \label{eq:LinearizedKinetic12}\\
\left\{v\cdot \partial + c(x, \Gamma)\right\}\varepsilon^3(x,\Gamma) + \omega_\mathrm{s}(E) \im S(x, \Gamma) = - c(x, \Gamma),
        \label{eq:LinearizedKinetic3}
\end{numcases}
\end{widetext}
up to linear order in $\varepsilon$.
The right-hand sides (r.h.s.) of Eqs.~(\ref{eq:LinearizedKinetic0})-(\ref{eq:LinearizedKinetic3}) are the source of the linear evolution, and the DR of this system of linear equations depends only on the left-hand sides (l.h.s.).
We can confirm that $\varepsilon^0$ is included only in Eq.~(\ref{eq:LinearizedKinetic0}) and is decoupled from the other components of $\varepsilon$.
Additionally, Eq.~(\ref{eq:LinearizedKinetic0}) gives the trivial DR for free-streaming massless particles:
\begin{align}
    v\cdot k = 0,
    \label{eq:DRForFree}
\end{align}
which has no instability.
Therefore, our main target is the DR for Eqs.~(\ref{eq:LinearizedKinetic12}) and (\ref{eq:LinearizedKinetic3}).

If we neglect flavor mixing, then from Eq.~(\ref{eq:Kinetic}), we have $c = v\cdot\partial \ln f_\mathrm{c} \sim \mathscr{C}/f_\mathrm{c}$, which is of the same order of magnitude as the inverse of the mean free path of the neutrinos.
This quantity is usually much smaller than the growth rate of neutrino flavor instabilities and can be safely neglected when computing the DR.
The values of $\omega_{\mathrm{c/s}}$ are also minimal compared to the growth rate for the typical energy of supernova neutrinos when we focus on a sufficiently small radius in a supernova.
As the radius increases and the neutrino density decreases, however, $\omega_{\mathrm{c/s}}$ becomes comparable to the growth rate, and a gradual transition to vacuum oscillation occurs.
The flavor instability when the $\omega_{\mathrm{c/s}}$ values are negligible is called a fast instability; when an instability with finite $\omega_{\mathrm{c/s}}$ vanishes in the limit of $\omega_{\mathrm{c/s}}\to 0$, it is called a slow instability~\cite{Airen2018}.
We note that Eqs.~(\ref{eq:LinearizedKinetic12}) and (\ref{eq:LinearizedKinetic3}) are coupled with each other via the terms proportional to $\omega_{\mathrm{s}}$.
In general, it is not appropriate to neglect $\omega_\mathrm{s}$ when focusing on slow instabilities, although \citet{Airen2018} did so.

Here, we focus on fast instabilities; we neglect $\omega_{\mathrm{c/s}}$ as well as $c$ on the l.h.s. of Eqs.~(\ref{eq:LinearizedKinetic12}) and (\ref{eq:LinearizedKinetic3}):
\begin{widetext}
\begin{numcases}{}
v\cdot \left\{i\partial -\Lambda_c(x)\right\}S(x,\Gamma) + \sqrt{2}G_F\int d\Gamma' f_\mathrm{c}(\Gamma')v\cdot v'S(x,\Gamma') = -\omega_\mathrm{s}(E),
        \label{eq:LinearizedKineticFast12}\\
v\cdot \partial\varepsilon^3(x,\Gamma) = -c(x, \Gamma).
        \label{eq:LinearizedKineticFast3}
\end{numcases}
\end{widetext}
These equations are no longer coupled with each other, and the DR for $\varepsilon^3$ is also given by Eq.~(\ref{eq:DRForFree}).
By taking the average over the energy $E$, Eq.~(\ref{eq:LinearizedKineticFast12}) is rewritten as
\begin{align}
    &v\cdot \left\{i\partial -\Lambda_c(x)\right\}\mathscr{S}(x,\vec{v})\nonumber\\
    &+ \int \dfrac{d^2\vec{v}'}{4\pi} \mathscr{G}(\vec{v}')v\cdot v'\mathscr{S}(x,\vec{v}') = \tilde{s}(x,\vec{v}),
    \label{eq:LinearizedKineticFast}
\end{align}
where
\begin{align}
    \mathscr{G}(\vec{v}) \equiv \sqrt{2}G_F\int_{-\infty}^\infty \dfrac{dE E^2}{2\pi^2}f_\mathrm{c}(\Gamma)
\end{align}
is the ELN angular distribution and
\begin{align}
    \mathscr{S}(x,\vec{v}) \equiv \int_{-\infty}^{\infty}\dfrac{dE E^2}{2\pi^2}\dfrac{f_\mathrm{c}(\Gamma)}{\mathscr{G}(\vec{v})}S(x,\Gamma).
\end{align}
Here, we omit a concrete expression for the source term $\tilde{s}$ and consider $\tilde{s}$ to depend on the spacetime position $x$ and the flight direction $\vec{v}$, although the r.h.s. of Eq.~(\ref{eq:LinearizedKineticFast12}) does not, for the reason discussed below.

Equation~(\ref{eq:seed}) can be separated into two contributions: (1) the homogeneous part $(0, 0, -\omega_s(E), 0)^\mathrm{T}$ and (2) the inhomogeneous part, which is the remainder.
The homogeneous part arises from the vacuum oscillation term and induces only modes with the wave vector $\vec{k} = 0$.
On the other hand, the inhomogeneous part depends on the position and can induce all modes.
This study focuses on the latter, and we consider the spatiotemporal behaviors of the linear response to this inhomogeneous source term.

We need to be careful when treating a fast instability.
When we omit $\omega_\mathrm{s}$ on the l.h.s. of Eqs.~(\ref{eq:LinearizedKinetic12}) and (\ref{eq:LinearizedKinetic3}), these equations are decoupled from each other, and the perturbation $S$ seems to be induced only by the homogeneous source $-\omega_\mathrm{s}$, as in Eq.~(\ref{eq:LinearizedKineticFast12}).
Even a small $\omega_\mathrm{s}$, however, can convert the inhomogeneous source $-c$ in Eq.~(\ref{eq:LinearizedKinetic3}) into $S$, and it then grows exponentially when a flavor eigenstate is unstable.
In other words, the effective source term proportional to $\omega_\mathrm{s}c$ should also be contained in Eq.~(\ref{eq:LinearizedKineticFast12}) in the fast regime to mimic the contribution from the inhomogeneous source term, and therefore, $\tilde{s}$ should be considered to have an $(x, \vec{v})$ dependency.

\subsection{Two-beam model}
To investigate the instabilities in detail, we adopt a 2-beam neutrino model in which the ELN angular distribution is expressed as
\begin{align}
    \mathscr{G}_{\vec{v}}=\mathscr{G}_{1}\delta(\vec{v},\vec{v}_{1})+\mathscr{G}_{2}\delta(\vec{v},\vec{v}_{2}),
\end{align}
where $\vec{v}_{1}$ and $\vec{v}_{2}$ are the directions of the neutrino beams and $\mathscr{G}_{1}$ and $\mathscr{G}_{2}$ correspond to their ELN intensities.
The delta function $\delta(\vec{v},\vec{v}')$ on the unit sphere is defined as
\begin{align}
    \int\frac{d^2\vec{v}'}{4\pi}\delta(\vec{v},\vec{v}')f(\vec{v}') = \int\frac{d^2\vec{v}'}{4\pi}f(\vec{v}')\delta(\vec{v}',\vec{v}) = f(\vec{v})
\end{align}
for an arbitrary continuous function $f$ on the unit sphere.
In addition, we focus on a sufficiently short time and a sufficiently small region such that the temporal and spatial variations in $\Lambda_\mathrm{c}(x)$ can be treated as constant.
In this model, Eq.~(\ref{eq:LinearizedKineticFast}) can be written as
\begin{align}
    \mat{D}(i\partial)\tilde{\mathscr{S}}(x) = \tilde{s}(x),
    \label{eq:LinearizedKinetic2Beam}
\end{align}
where $\mat{D}$ and $\tilde{\mathscr{S}}$ are defined as
\begin{align}
    \mat{D}(i\partial) \equiv
    \begin{pmatrix}
        v_{1}\cdot (i\partial) & v_{1}\cdot v_{2}\mathscr{G}_{2}\\
        v_{1}\cdot v_{2}\mathscr{G}_{1} & v_{2} \cdot (i\partial)
    \end{pmatrix}
\end{align}
and
\begin{align}
    \tilde{\mathscr{S}}(x) \equiv e^{i\Lambda_\mathrm{c}\cdot x}
    \begin{pmatrix}
        \mathscr{S}(x,\vec{v_{1}})\\
        \mathscr{S}(x,\vec{v_{2}})
    \end{pmatrix}.
    \label{eq:DefOfStilde}
\end{align}
We note that $\Lambda_\mathrm{c}$ is a real 4-vector and does not affect the instability.
For instability analysis, the DR given by the zeros of
\begin{align}
    \Delta(k) \equiv \det \mat{D}(k) = (v_1\cdot k)(v_2\cdot k)- \epsilon
\end{align}
is important, where
\begin{align}
    \epsilon \equiv (v_1\cdot v_2)^2\mathscr{G}_1\mathscr{G}_2.
\end{align}

\section{Spatiotemporal instability analysis}

\subsection{General formulation}
The spatiotemporal behaviors of the perturbation $\tilde{\mathscr{S}}$ are obtained by solving Eq.~(\ref{eq:LinearizedKinetic2Beam}) for the appropriate initial and boundary conditions.
In our setup, the inhomogeneous source $\tilde{s}$, which takes the form of a wave packet in realistic situations, drives the perturbation.
To treat such perturbations, we simply focus on the Green's function $\mat{G}$ for the linearized equation defined by
\begin{align}
    \mat{D}(i\partial)\mat{G}(x) = \delta^4(x)\mat{I}_2
\end{align}
because its asymptotic behavior at $t\to\infty$ is essentially the same as that of the perturbations~\cite{Briggs1964,Landau1997}.

Here, we discuss the necessity of this prescription.
\citet{Izaguirre2017} stated that a complex wave vector $\vec{k}$ for a real $\omega$ in a DR implies ``spatial instability'', and several studies based on linear analysis for stationary solutions have actually investigated this kind of ``instability''.
We should, however, be careful about what this really means~\cite{Yi2019}.
The existence of a complex $\vec{k}$ for a real $\omega$ ensures only the existence of spatially exponentially growing/decaying modes that oscillate or are constant in time.
Therefore, such a ``spatial instability'' seems to appear if we consider stationary solutions of Eq.~(\ref{eq:Kinetic}).
However, it is not guaranteed that stationary solutions, and hence this ``spatial instability'', will be realized after time evolution.
To ensure this, we should at least investigate the time evolution of the perturbations from the stationary solution and confirm the damping of the perturbations.
Moreover, even spatial growth is not guaranteed by this ``spatial instability''.
For example, if we consider the diffusion equation $(\partial_t - \partial_x^2)f(t,x) = 0$, which has a ``spatial instability'' from the DR $k = \pm\sqrt{i\omega}$, a constant source imposed at a certain point decays in space.
After all, such a ``spatial instability'' is not always relevant in realistic situations.
On the other hand, a complex $\omega$ for a real $\vec{k}$ does imply instability in time evolution.
The explicit form of the DRs $\omega(\vec{k})$ for $\vec{k}\in\mathbb{R}^3$ directly shows the time evolution of each plane wave mode.
However, how perturbations in the form of wave packets, which are superpositions of uncountably infinite numbers of plane waves, actually behave in spacetime cannot be trivially determined from the DRs; hence, we should focus on the behaviors of the Green's function, which is also a superposition of all modes.

The asymptotic behavior of $\mat{G}$ can be evaluated by using the Lefschetz thimble method~\cite{Morinaga2020}.
By applying the Laplace transform for time and the Fourier transform for space, $\mat{G}$ is expressed as
\begin{align}
    \mat{G}(t,\vec{x}+\vec{u}t) = \int_{\mathcal{M}}\dfrac{d^4k}{(2\pi)^{4}}e^{-ik\cdot ut}e^{i\vec{k}\cdot\vec{x}}\mat{D}(k)^{-1},
    \label{eq:GIntM}
\end{align}
where $\mathcal{M}\equiv L\times\mathbb{R}^3$ is the Laplace-Fourier contour and $(u^\mu)\equiv(1,\vec{u})$ with the parameter vector $\vec{u}\in\mathbb{R}^3$.
The residue theorem yields
\begin{align}
    \mat{G}(t,\vec{x}+\vec{u}t) = \dfrac{\theta(t)}{(2\pi)^d i}\int_{\mathcal{C}}d^d\vec{k}\dfrac{e^{-ik\cdot ut}e^{i\vec{k}\cdot\vec{x}}}{\partial_0\Delta(k)}\adj\mat{D}(k),
    \label{eq:GintC}
\end{align}
where $\mathcal{C}\equiv\mathcal{D}\cap\left(\mathbb{C}\times\mathbb{R}^d\right)$ and $\mathcal{D}\equiv\left\{k\in\mathbb{C}^{d+1}|\Delta(k)=0\right\}$.

The contour of integral $\mathcal{C}$ can be deformed to a sum of Lefschetz thimbles $\{\mathcal{J}_\sigma\}$ as follows:
\begin{align}
    \mathcal{C} \cong \sum_{\sigma}\braket{\mathcal{C},\mathcal{K}_\sigma}\mathcal{J}_\sigma,
\end{align}
where $\braket{\mathcal{C},\mathcal{K}_\sigma}$ is the intersection number of $\mathcal{C}$ and $\mathcal{K}_\sigma$ and the $\{\mathcal{K}_\sigma\}$ are called the dual thimbles.
The Lefschetz thimble $\mathcal{J}_\sigma$ and the dual thimble $\mathcal{K}_\sigma$ are associated with the critical points of the ``height function'' defined on $\mathcal{D}$ as follows:
\begin{align}
    h(k)\equiv \re (-ik\cdot u) = \im k\cdot u,
\end{align}
which corresponds to the real part of the exponent of the dominant factor $e^{-ik\cdot u}$.
The critical points are given by the stationary conditions for $h$ constrained on $\mathcal{D}$, which can be simplified as
\begin{align}
    \begin{cases}
        \Delta (k_\sigma) = 0,\\
        \left(\partial_i-u_i\partial_0\right)\Delta(k_\sigma) = 0.
    \end{cases}
\end{align}
The Lefschetz thimble associated with the critical point $k_\sigma$ is obtained by solving the upward flow equation
\begin{align}
    \dfrac{dK^\alpha(s)}{ds} = iu_\beta\left[\delta^{\beta\alpha}-\delta^{\beta\gamma}\dfrac{\partial_\gamma\Delta\overline{\partial_\delta\Delta}}{\|\partial\Delta\|^2}\delta^{\delta\alpha}\right]_{k=K(s)}
\end{align}
with the boundary condition $\lim_{s\to\infty}K(s) = k_\sigma$, and the dual thimble $K_\sigma$ is found by solving the same differential equation with the boundary condition $\lim_{s\to-\infty}K(s) = k_\sigma$.

When $|\vec{k}_\sigma\cdot\vec{x}| \ll |k_\sigma\cdot ut|$, the integral over the Lefschetz thimble $\mathcal{J}_\sigma$ is dominated by the contribution around the critical point $k_\sigma$, so the integral is evaluated as follows:
\begin{align}
    \mat{G}(\vec{x}+\vec{u}t) \sim \dfrac{1}{t^{d/2}}e^{-ik_{\mathrm{m}}\cdot ut}\adj\mat{D}(k_{\mathrm{m}}),
\end{align}
where $k_{\mathrm{m}}$ is the critical point with the maximum height $h(k_{\mathrm{m}})$ among all critical points satisfying $\braket{\mathcal{C},\mathcal{K}_\sigma}\neq 0$.

Now, we can classify instabilities in terms of the behavior of $\mat{G}$.
If $\mat{G}$ grows for $\vec{u} = \vec{0}$, this means that perturbations will grow at every point throughout the whole space; such an instability is called an absolute instability.
On the other hand, if $\mat{G}$ decays for $\vec{u} = \vec{0}$ but grows for another $\vec{u}$, then the perturbations will grow but flow away from the generated point; such an instability is called a convective instability.
If $\mat{G}$ decays for all $\vec{u}$, then the system is stable.

\subsection{Two-dimensional perturbation}
Now, we can conduct a spatiotemporal instability analysis of the 2-beam model considering 4-dimensional perturbations.
In this subsection, however, we first perform a 2-dimensional analysis for two reasons: we would like to reveal the problems with the 2-dimensional analysis, and in the case of the 2-beam model, the results of the 2-dimensional analysis are also helpful for the 4-dimensional analysis.

For later use, we introduce some vector notations as follows:
\begin{itemize}
\item Normalized vector:
\begin{align}
        \hat{\vec{a}} \equiv \dfrac{\vec{a}}{|\vec{a}|}
    \end{align}
\item Vector component parallel to $\vec{b}$:
\begin{align}
        a_{\vec{b}} \equiv \vec{a}\cdot\hat{\vec{b}}
    \end{align}
\item Projection of a vector onto the direction parallel to $\vec{b}$:
\begin{align}
        \vec{a}_{\vec{b}} \equiv a_{\vec{b}}\hat{\vec{b}}
    \end{align}
\item Projection of a vector onto the plane perpendicular to $\vec{b}$:
\begin{align}
        \vec{a}_{\perp\vec{b}} \equiv \vec{a} - \vec{a}_{\vec{b}}.
    \end{align}
\end{itemize}
In addition, we define
\begin{align}
    \begin{cases}
        \vec{V}\equiv\dfrac{\vec{v}_{1}+\vec{v}_{2}}{2},\\
        \vec{v}\equiv\dfrac{\vec{v}_{1}-\vec{v}_{2}}{2},
    \end{cases}
\end{align}
which are perpendicular to each other because $|\vec{v}_1|=|\vec{v}_2|=1$.

The 2-dimensional Green's function $\mat{G}^{(2)}_{\vec{n}}$ can be defined as
\begin{align}
    \mat{D}(i\partial)\mat{G}^{(2)}_{\vec{n}}(t,x_{\vec{n}}) = \delta(t)\delta(x_{\vec{n}})\mat{I}_2,
\end{align}
where the unit vector $\vec{n}$ gives the direction along which we consider perturbations and $x_{\vec{n}} \equiv \vec{x}\cdot\vec{n}$ as defined above.
When we take $\vec{n}$ to be the $z$-direction, the situation is the same as in Ref.~\cite{Capozzi2017}.
The Laplace-Fourier transform for time and the $\vec{n}$-direction yields
\begin{align}
    \mat{D}(\omega,k\vec{n})\mat{G}^{(2)}_{\vec{n}}(\omega,k) = \mat{I}_2,
\end{align}
and $\mat{G}^{(2)}_{\vec{n}}$ can be expressed as
\begin{align}
    \mat{G}^{(2)}_{\vec{n}}(t,x+ut) =& \int_{L}\dfrac{d\omega}{2\pi}\int_{-\infty}^{\infty}\dfrac{dk}{2\pi}e^{-i(\omega -ku)t}e^{ikx}\mat{D}(\omega,k\vec{n})^{-1}.
    \label{eq:G2Int}
\end{align}
Although we should compute the asymptotic behaviors of $\mat{G}^{(2)}_{\vec{n}}(t,x+ut)$ for all $u$ using the method given above, we can achieve an equivalent task by considering only $u=0$ in this case.
When the variables of integration are transformed such that
\begin{align}
    \begin{pmatrix}
        w\\
        \kappa
    \end{pmatrix}
    =
    \begin{pmatrix}
        1 & -u\\
        0 & u - V_{\vec{n}}
    \end{pmatrix}
    \begin{pmatrix}
        \omega\\
        k
    \end{pmatrix},
\end{align}
Eq.~(\ref{eq:G2Int}) becomes
\begin{align}
    \mat{G}^{(2)}_{\vec{n}}(t,x+ut) =& \int_{L}\dfrac{dw}{2\pi}\int_{-\infty}^{\infty}\dfrac{d\kappa}{2\pi|u-V_{\vec{n}}|}e^{-iwt}e^{i\frac{\kappa}{u-V_{\vec{n}}}x}\nonumber\\
    &\times\mat{D}\left(w+\dfrac{u\kappa}{u-V_{\vec{n}}},\dfrac{\kappa}{u-V_{\vec{n}}}\vec{n}\right)^{-1}.
\end{align}
This integral can be evaluated by means of the Lefschetz thimble method with height function $h(w,\kappa)\equiv\im w$ and DR
\begin{align}
    \tilde{\Delta}(w,\kappa) \equiv& \det\mat{D}\left(w+\dfrac{u\kappa}{u-V_{\vec{n}}},\dfrac{\kappa}{u-V_{\vec{n}}}\vec{n}\right)\nonumber\\
    =& w^2 + 2\kappa w + (1-\alpha^2)\kappa^2 - \epsilon,
    \label{eq:DRFor2Beam}
\end{align}
where
\begin{align}
    \alpha(u,\vec{n}) \equiv \dfrac{v_{\vec{n}}}{u-V_{\vec{n}}}.
\end{align}
The critical points for this system are given by
\begin{align}
    \begin{cases}
        \tilde{\Delta} (w,\kappa) = 0,\\
        \dfrac{\partial}{\partial \kappa}\tilde{\Delta}(w,\kappa) = 0.
    \end{cases}
\end{align}
This system of equations can be solved analytically, and we obtain two critical points $(w_+,\kappa_+)$ and $(w_-,\kappa_-)$:
\begin{align}
    (w_\pm,\kappa_\pm) = \left(\pm\dfrac{\sqrt{(\alpha^2-1)\epsilon}}{|\alpha|}, \mp\dfrac{1}{|\alpha|}\sqrt{\dfrac{\epsilon}{\alpha^2-1}}\right).
\end{align}
The dual thimbles $\mathcal{K}_{\pm}$ can be obtained by solving the upward flow equation
\begin{align}
    \dfrac{dK^\alpha(s)}{ds} = i\left[\delta^{0\alpha}-\delta^{0\gamma}\dfrac{\partial_\gamma\tilde{\Delta}\overline{\partial_\delta\tilde{\Delta}}}{\|\partial\tilde{\Delta}\|^2}\delta^{\delta\alpha}\right]_{k=K(s)}
\end{align}
with the boundary condition $\lim_{s\to-\infty}K(s) = k_\sigma$.
Numerically, this is achieved by modifying the boundary condition as follows: $K(0)=k_{\pm}+(0,H_\pm\delta)$, where $\delta\in\mathbb{R}$ is a small quantity and
\begin{align}
    H_\pm \equiv \left.\dfrac{\partial_\kappa^2\tilde{\Delta}}{\partial_w\tilde{\Delta}}\right|_{(w,\kappa)=(w_\pm,\kappa_\pm)}.
\end{align}

\begin{figure}[htb]
    \includegraphics[width=0.8\linewidth]{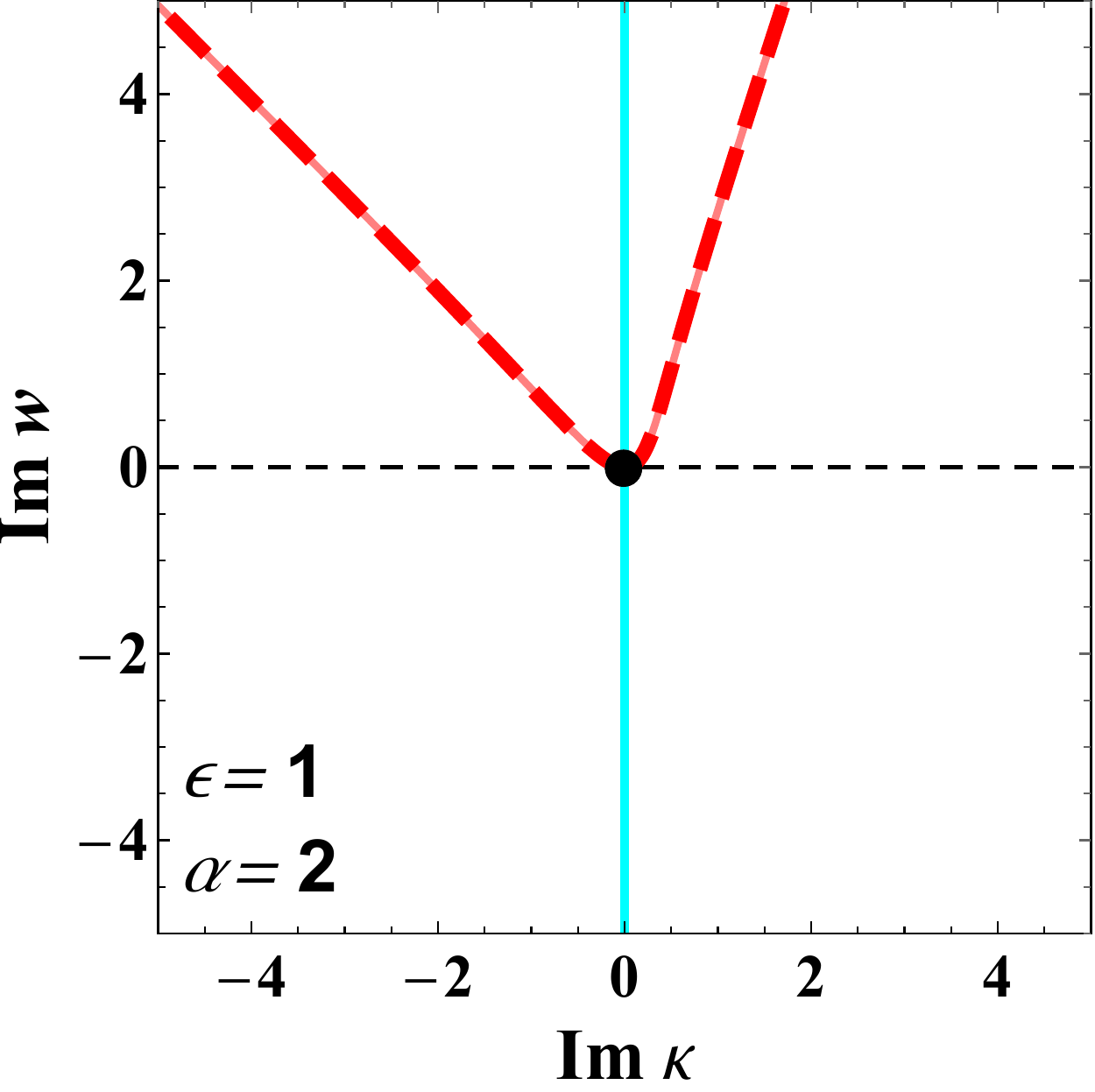}
    \includegraphics[width=0.8\linewidth]{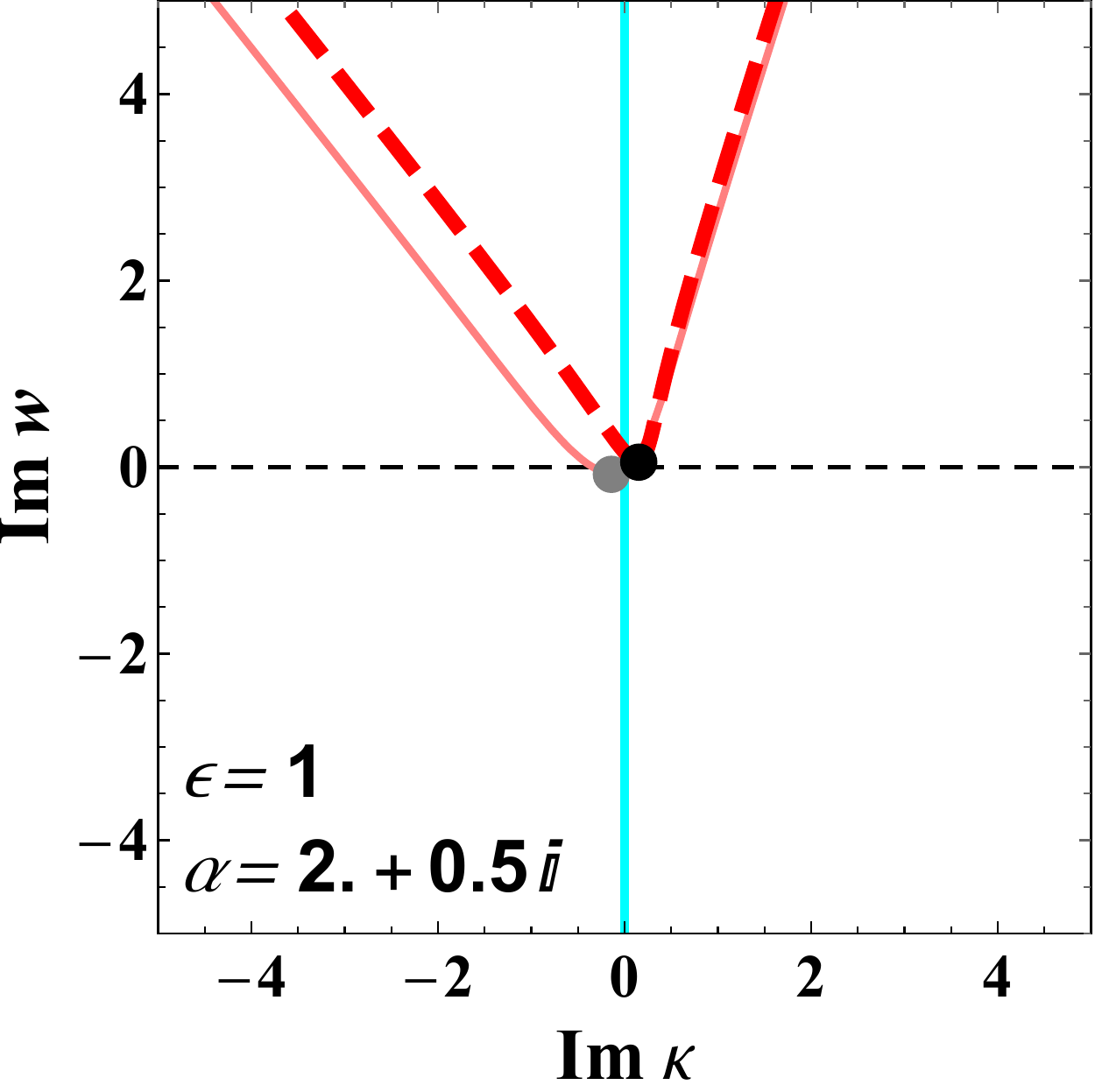}
\caption{The critical points (a black point for $k_+$ and a gray point for $k_-$) and dual thimbles (a red dashed line for $\mathcal{K}_+$ and a pink solid line for $\mathcal{K}_-$) for the DR $\tilde{\Delta}(w,\kappa)=0$ with $\epsilon = 1$ and $\alpha = 2$. The cyan line represents $\im \kappa = 0$, part of which corresponds to $\mathcal{C}$. The bottom panel shows the same figure with a small imaginary part added to $\alpha$ for readability.}
    \label{fig:DualThimbles_e1a2}
\end{figure}

\begin{figure}[htb]
    \includegraphics[width=0.8\linewidth]{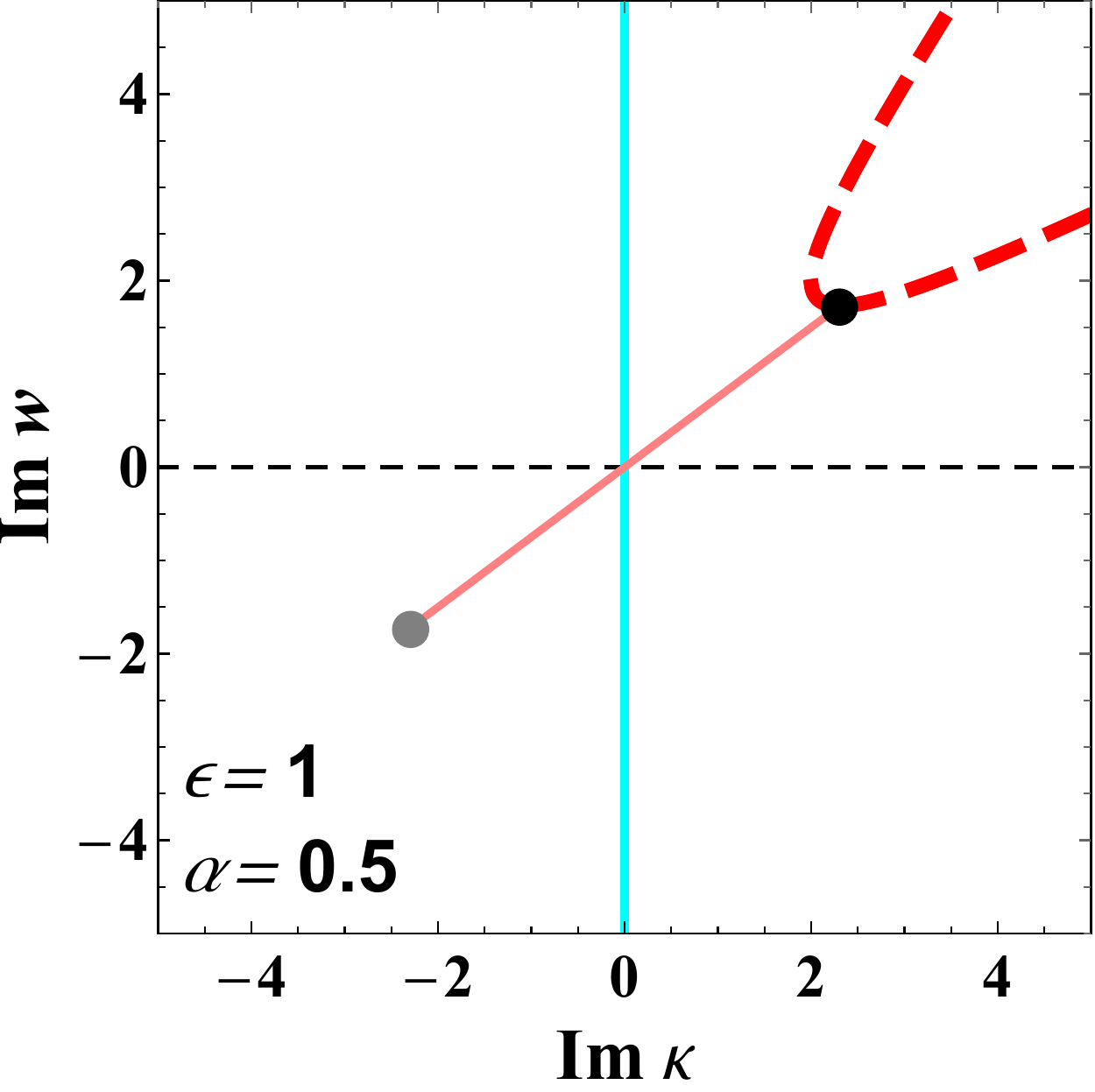}
    \includegraphics[width=0.8\linewidth]{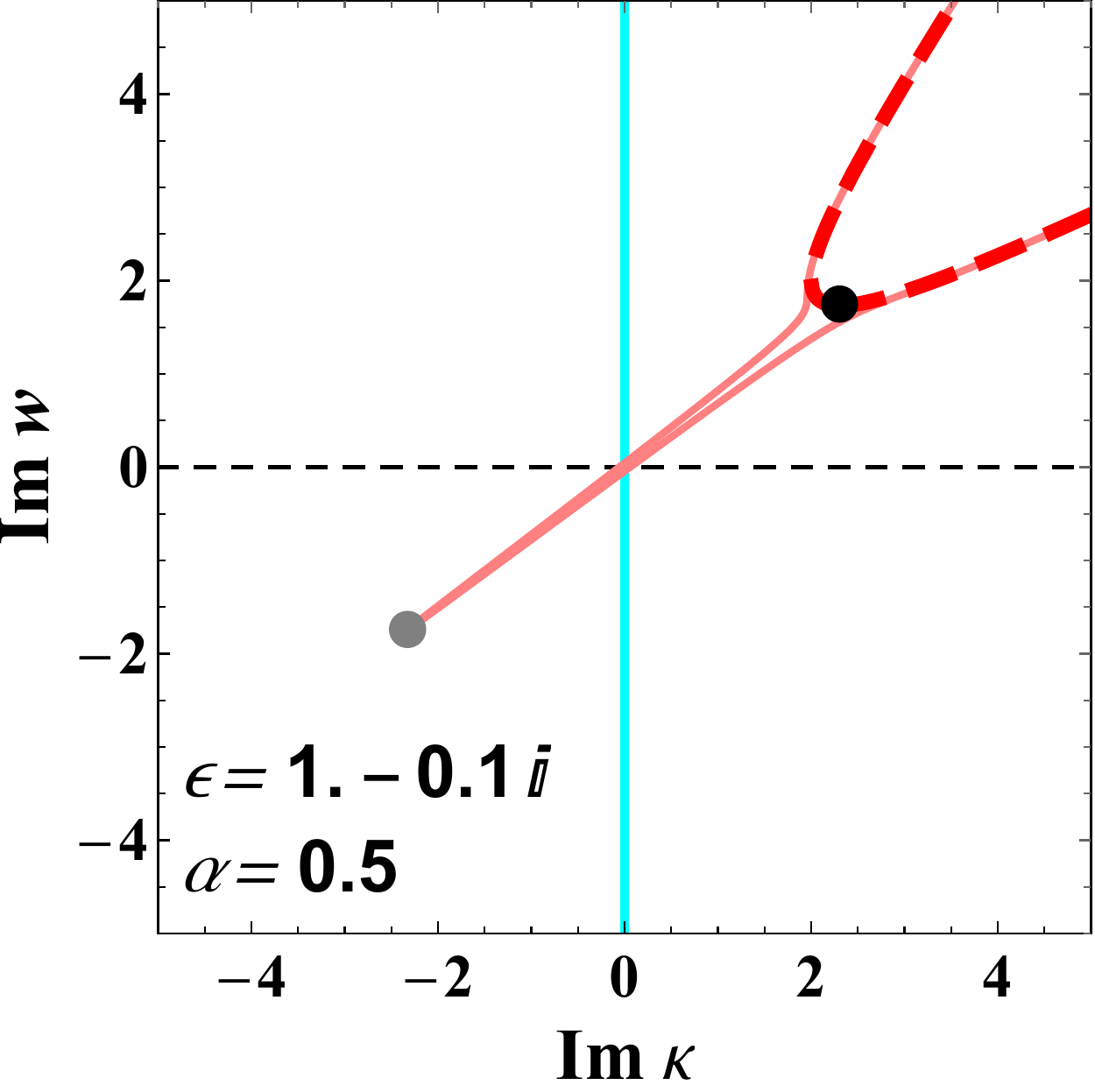}
\caption{The same figure as Fig.~\ref{fig:DualThimbles_e1a2} for $\epsilon=1$ and $\alpha=0.5$. The bottom panel shows the same figure with a small imaginary part added to $\epsilon$.}
    \label{fig:DualThimbles_e1a05}
\end{figure}

\begin{figure}[htb]
    \includegraphics[width=0.8\linewidth]{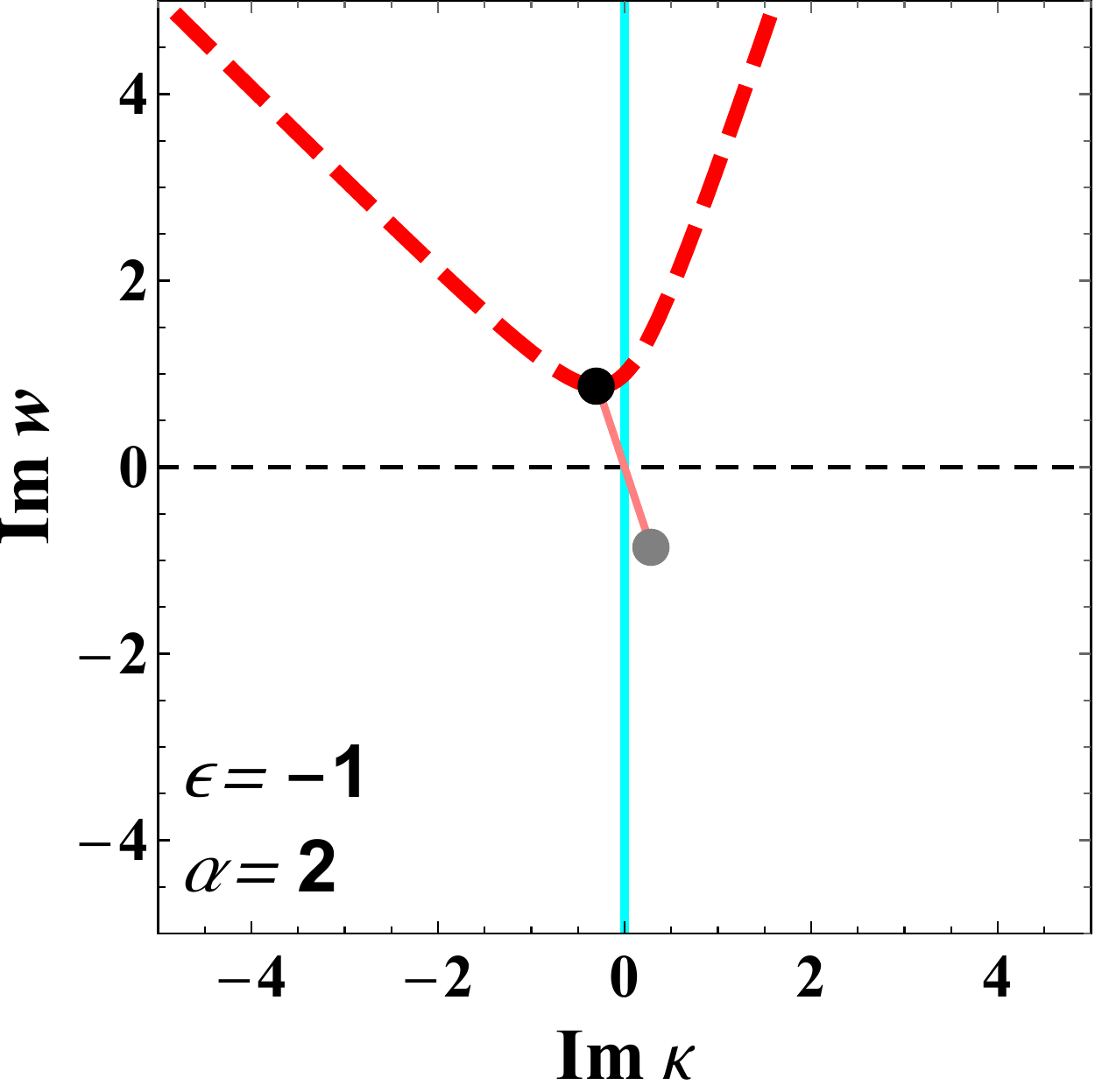}
    \includegraphics[width=0.8\linewidth]{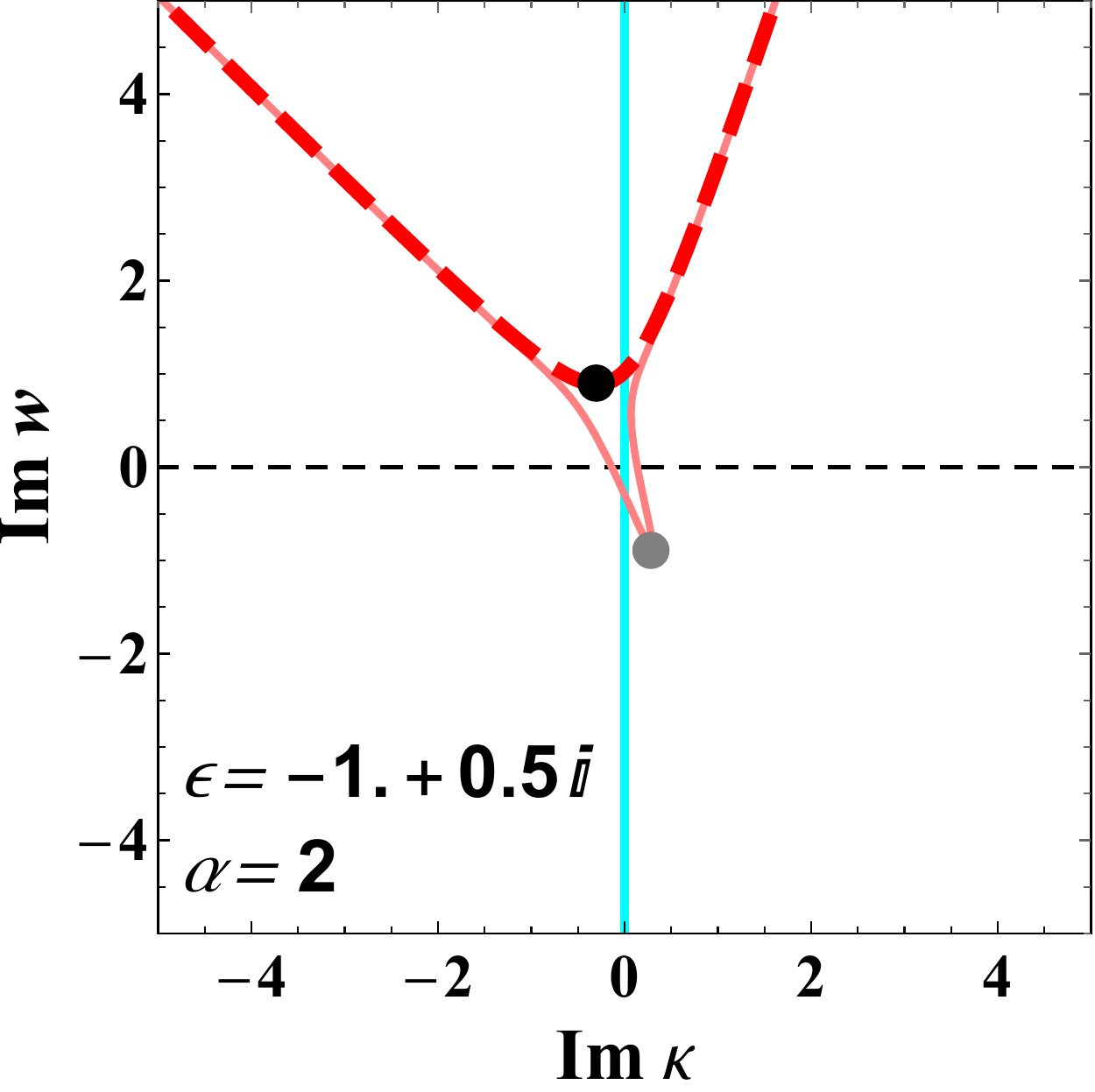}
\caption{The same figure as Fig.~\ref{fig:DualThimbles_e1a2} for $\epsilon=-1$ and $\alpha=2$. The bottom panel shows the same figure with a small imaginary part added to $\epsilon$.}
    \label{fig:DualThimbles_em1a2}
\end{figure}

\begin{figure}[htb]
    \includegraphics[width=0.8\linewidth]{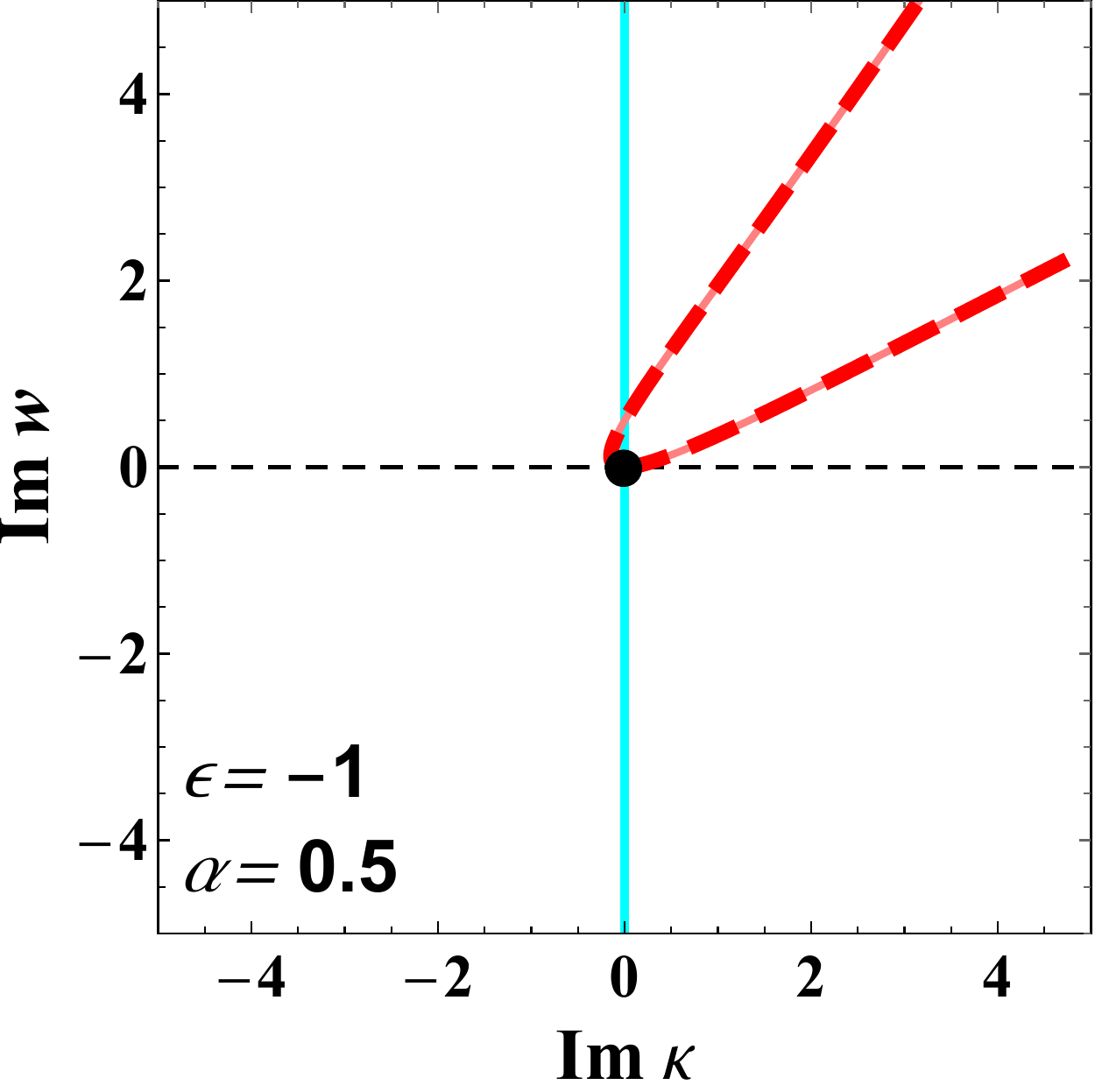}
    \includegraphics[width=0.8\linewidth]{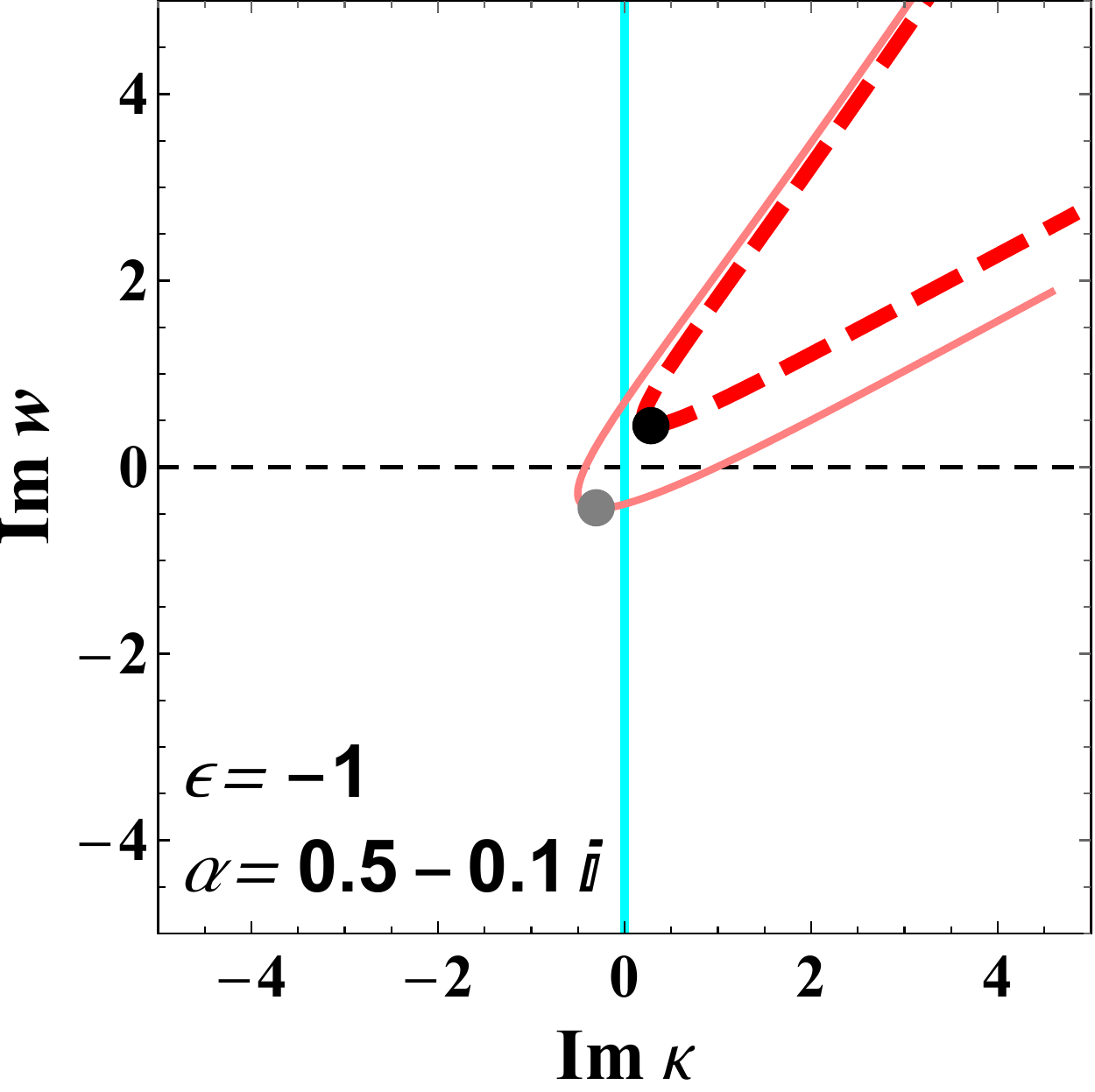}
\caption{The same figure as Fig.~\ref{fig:DualThimbles_e1a2} for $\epsilon=-1$ and $\alpha=0.5$. The bottom panel shows the same figure with a small imaginary part added to $\alpha$.}
    \label{fig:DualThimbles_em1a05}
\end{figure}

The critical points and dual thimbles are shown in Figs.~\ref{fig:DualThimbles_e1a2}, \ref{fig:DualThimbles_e1a05}, \ref{fig:DualThimbles_em1a2} and \ref{fig:DualThimbles_em1a05} for several values of $\epsilon$ and $\alpha$.
All of these figures show the projections onto the $\im \kappa-\im\omega$ plane, and the original contour $\mathcal{C}$ lies within the profile of $\im \kappa =0$ (cyan line).
What we need to pay attention to is the intersection numbers of $\mathcal{C}$ and $\mathcal{K}_\pm$.
For example, in Fig.~\ref{fig:DualThimbles_e1a2}, each of the dual thimbles $\mathcal{K}_\pm$ has a single intersection with $\mathcal{C}$, i.e., $|\braket{\mathcal{C},\mathcal{K}_\pm}|=1$, which means that the Lefschetz thimbles $\mathcal{J}_\pm$ both contribute to the integral.
We note that the sign of the intersection number $\braket{\mathcal{C}, \mathcal{K}_\sigma}$ can be determined by giving $\mathcal{K}_\sigma$, and consequently $\mathcal{J}_\sigma$, an orientation.
For our purposes, however, factors other than the dominant exponential factor are not important, so we do not consider them.
In contrast, the dual thimbles in Fig.~\ref{fig:DualThimbles_em1a05} seem to intersect with $\mathcal{C}$ twice.
In fact, however, they cross $\mathcal{C}$ from opposite sides at each intersection, and the intersection numbers for the two intersections are $+1$ and $-1$; consequently, the sum is $\braket{\mathcal{C},\mathcal{K}_\pm}=0$.


The situation is more complicated in Figs.~\ref{fig:DualThimbles_e1a05} and \ref{fig:DualThimbles_em1a2}.
The dual thimble $\mathcal{K}_-$ is absorbed into the higher critical point $(w_+,\kappa_+)$ and cannot be well defined.
This situation is understood to be related to the fact that these parameter sets lie on Stokes rays \cite{Witten2010}, and we can avoid it by adding small perturbations to the imaginary parts of the parameters (see the bottom panels in the figures).
In Fig.~\ref{fig:DualThimbles_em1a2}, it can be confirmed in the bottom figure that both of the dual thimbles have intersections with $\mathcal{C}$, i.e., $|\braket{\mathcal{C},\mathcal{K}_\pm}|=1$.
In Fig.~\ref{fig:DualThimbles_e1a05}, the dual thimble $\mathcal{K}_+$ apparently does not intersect with $\mathcal{C}$, and $\mathcal{K}_-$ also does not, in the same manner as in Fig.~\ref{fig:DualThimbles_em1a05}; thus, neither of the Lefschetz thimbles influences the integral.
It should be noted that the signs of $\braket{\mathcal{C},\mathcal{K}_-}$ can be either positive or negative, depending on the perturbations added. However, this is not important for the asymptotic behavior of the Green's functions because they are not the highest critical points to which the associated Lefschetz thimble contributes to the integral.

Although we show figures only for these 4 parameter sets, we can confirm that the intersection numbers depend only on the signs of $\epsilon$ and $|\alpha|-1$.
Therefore, for $|\alpha|<1$, the Green's function $\mat{G}^{(2)}_{\vec{n}}(t,\vec{u}t)$ is exactly $\mat{0}$ for all $t$.
For $|\alpha|>1$ and $\epsilon>0$, both of the Lefschetz thimbles $\mathcal{J}_\pm$ contribute to the integral, but the growth rate $\im w_\pm = 0$ means that the Green's function is damped as $\sim 1/\sqrt{t}$.
For $|\alpha|>1$ and $\epsilon<0$, the growth rate is $\im w_+ = \sigma^{(2)}$, where
\begin{align}
    \sigma^{(2)}(u,\vec{n}) \equiv \sqrt{-\epsilon\left(1-\dfrac{1}{\alpha(u,\vec{n})^2}\right)}.
\end{align}
As a result, the asymptotic behavior of $\mat{G}^{(2)}_{\vec{n}}$ can be summarized as
\begin{align}
    &\mat{G}^{(2)}_{\vec{n}}(t,x+ut) \nonumber\\
    &\sim
    \begin{cases}
        \dfrac{1}{\sqrt{t}}e^{\sigma^{(2)} t}\mat{M}^{(2)} & (\epsilon<0,\ |u-V_{\vec{n}}|<|v_{\vec{n}}|),\\
        \mat{0} & (\mathrm{otherwise}),
    \end{cases}
    \label{eq:G2}
\end{align}
where $\mat{M}^{(2)}$, whose concrete expression is omitted, is a constant matrix independent of $t$.

Based on these results, we can classify the instability for each parameter set as shown in Table~\ref{tab:InstabilityCondition2D}.
For $\epsilon\geq 0$, the Green's function decays in time, meaning that the flavor eigenstate is stable.
If $\epsilon<0$ and $|V_{\vec{n}}|<|v_{\vec{n}}|$ are satisfied, the Green's function grows in the rest frame with $u=0$, corresponding to an absolute instability.
Finally, if $\epsilon<0$ is satisfied and $|V_{\vec{n}}|<|v_{\vec{n}}|$ is not, then the Green's function decays for $u=0$ but grows for $u\in(V_{\vec{n}}-|v_{\vec{n}}|,V_{\vec{n}}+|v_{\vec{n}}|)$, corresponding to a convective instability.
\begin{table}[htb]
    \caption{Associations between instabilities and model parameters for 2-dimensional perturbations.}
    \begin{ruledtabular}
        \begin{tabular}{cc}
            Classification & Condition\\
            \hline
            Stable & $\epsilon\geq 0$ \\
            Convectively unstable & $\epsilon<0,\ |V_{\vec{n}}|\geq|v_{\vec{n}}|$\\
            Absolutely unstable & $\epsilon<0,\ |V_{\vec{n}}|<|v_{\vec{n}}|$ \\
        \end{tabular}
    \end{ruledtabular}
    \label{tab:InstabilityCondition2D}
\end{table}

The results are the same as those in Ref.~\cite{Capozzi2017} if we choose the $z$-direction as $\vec{n}$.
We stress, however, that in a 2-dimensional treatment, not only is the space of the perturbations restricted, but the interpretation of the results in 4-dimensional spacetime is also ambiguous.
The results depend on the direction $\vec{n}$, and in fact, it can happen that the system is convectively unstable for one $\vec{n}$ but absolutely unstable for another $\vec{n}$.
Needless to say, the nature of the true instability should not depend on which direction we choose as $\vec{n}$.
What we have done here is to consider a perturbation that takes the form of a wave packet in the direction parallel to $\vec{n}$ but is homogeneous in the directions perpendicular to $\vec{n}$ and investigate the corresponding spatiotemporal evolution.
Therefore, even if we were to investigate all possible directions $\vec{n}$, perturbations with the form of wave packets in 4-dimensional spacetime would not be taken into account.
How perturbations truly behave in 4-dimensional spacetime cannot be captured by a 2-dimensional analysis.
Nevertheless, a 2-dimensional analysis may be justified for systems that are open in the $\vec{n}$ direction but bounded in the directions perpendicular to $\vec{n}$.
Even in such a case, however, we need to conduct the analysis for all plane wave modes perpendicular to $\vec{n}$ with wave numbers $\vec{K}_{\perp\vec{n}}$, and the DR will be modified to
\begin{align}
    \Delta(\omega,k)=\det\mat{D}(\omega,k\vec{n}+\vec{K}_{\perp\vec{n}})=0
\end{align}
in general, although we treat only $\vec{K}_{\perp\vec{n}}=\vec{0}$ in the discussion above.

\subsection{Four-dimensional perturbation}
The general methodology for considering a 4-dimensional perturbation has already been presented in Sect.~III.~A.
In the 2-beam model, the integral in the calculation of the Green's function can be performed partially analytically, and the analysis reduces to the one in 2-dimensional spacetime.

Initially, $\mat{D}$ satisfies
\begin{align}
    \mat{D}(k) =& 
    \begin{pmatrix}
        \omega-|\vec{v}|k_{\vec{v}}-\vec{V}\cdot\vec{k}_{\perp\vec{v}} & v_1\cdot v_2\mathscr{G}_{2}\\
        v_1\cdot v_2\mathscr{G}_{1} & \omega+|\vec{v}|k_{\vec{v}}-\vec{V}\cdot\vec{k}_{\perp\vec{v}}
    \end{pmatrix}\nonumber\\
    =& \mat{D}(\omega-\vec{V}\cdot\vec{k}_{\perp\vec{v}},\vec{k}_{\vec{v}}),
    \label{eq:FormulaForD}
\end{align}
and the inverse matrix is expressed as
\begin{align}
    \mat{D}(k)^{-1}=\dfrac{1}{\Delta(k)}
    \begin{pmatrix}
        v_{2}\cdot k & -v_{1}\cdot v_{2}\mathscr{G}_{2}\\
        -v_{1}\cdot v_{2}\mathscr{G}_{1} & v_{1}\cdot k
    \end{pmatrix}.
\end{align}
The integral expression for $\mat{G}$ is given by Eq. (\ref{eq:GIntM}).
When we decompose the variables of the integral $\vec{k}$ into the components parallel and perpendicular to $\vec{v}$ and use Eq.~(\ref{eq:FormulaForD}), $\mat{G}$ is written as
\begin{align}
    \mat{G}(t,\vec{x}+\vec{u}t) =& \int_{L}\dfrac{d\omega}{2\pi}\int_{-\infty}^{\infty}\dfrac{dk_{\vec{v}}}{2\pi}\int_{\mathbb{R}^2}\dfrac{d^2\vec{k}_{\perp\vec{v}}}{2\pi}\nonumber\\
    &\times e^{-i(\omega-k_{\vec{v}}u_{\vec{v}})t}e^{ik_{\vec{v}}x_{\vec{v}}}e^{i\vec{k}_{\perp\vec{v}}\cdot (\vec{u}_{\perp\vec{v}}t+\vec{x}_{\perp\vec{v}})}\nonumber\\
    &\times \mat{D}(\omega-\vec{V}\cdot\vec{k}_{\perp\vec{v}},\vec{k}_{\vec{v}})^{-1}.
\end{align}
The variable transformation $\omega'\equiv\omega-\vec{V}\cdot\vec{k}_{\perp\vec{v}}$ yields
\begin{align}
    \mat{G}(t,\vec{x}+\vec{u}t) =& \int_{\mathbb{R}^2}\dfrac{d^2\vec{k}_{\perp\vec{v}}}{2\pi}e^{i\vec{k}_{\perp\vec{v}}\cdot [(\vec{u}_{\perp\vec{v}}-\vec{V})t+\vec{x}_{\perp\vec{v}}]}\nonumber\\
    &\times\int_{L}\dfrac{d\omega'}{2\pi}\int_{-\infty}^{\infty}\dfrac{dk_{\vec{v}}}{2\pi} e^{-i\omega't}e^{ik_{\vec{v}}(x_{\vec{v}}+u_{\vec{v}}t)}\nonumber\\
    &\times \mat{D}(\omega',\vec{k}_{\vec{v}})^{-1}.
\end{align}
Now, we note that the integral over $\vec{k}_{\perp\vec{v}}$ can be performed, and those over $\omega'$ and $k_{\vec{v}}$ can be replaced by the 2-dimensional Green's function $\mat{G}^{(2)}_{\hat{\vec{v}}}$:
\begin{align}
    \mat{G}(t,\vec{x}+\vec{u}t) = \delta^2([\vec{x}-(\vec{V}-\vec{u})t]_{\perp\vec{v}})\mat{G}_{\hat{\vec{v}}}^{(2)}\left(t,(\vec{x}+\vec{u}t)_{\vec{v}}\right).
\end{align}
Therefore, the DR to be considered is again the one expressed in Eq.~(\ref{eq:DRFor2Beam}), but $\alpha$ is now given by
\begin{align}
    \alpha(u_{\vec{v}},\hat{\vec{v}}) = \dfrac{|\vec{v}|}{u_{\vec{v}}}.
\end{align}
We can obtain the asymptotic behavior of $\mat{G}^{(2)}_{\hat{\vec{v}}}$ by substituting $\hat{\vec{v}}$ into $\vec{n}$ in Eq.~(\ref{eq:G2}), and the final result is
\begin{widetext}
\begin{align}
   \mat{G}(t,\vec{x}+\vec{u}t) \sim & 
    \begin{cases}
        \delta^{2}\left([\vec{x}-(\vec{V}-\vec{u})t]_{\perp\vec{v}}\right)\dfrac{1}{\sqrt{t}}e^{\sigma t}\mat{M} & (\epsilon<0,\ |u_{\vec{v}}|<|\vec{v}|),\\
        \mat{0} & (\mathrm{otherwise}),
    \end{cases}
    \label{eq:G}
\end{align}
\end{widetext}
where
\begin{align}
    \sigma(u_{\vec{v}}) \equiv \sqrt{-\epsilon\left\{1-\left(\dfrac{u_{\vec{v}}}{|\vec{v}|}\right)^2\right\}}
\end{align}
and $\mat{M}$ is a constant matrix, whose concrete expression is omitted.

\begin{figure}[htb]
    \includegraphics[width=\linewidth]{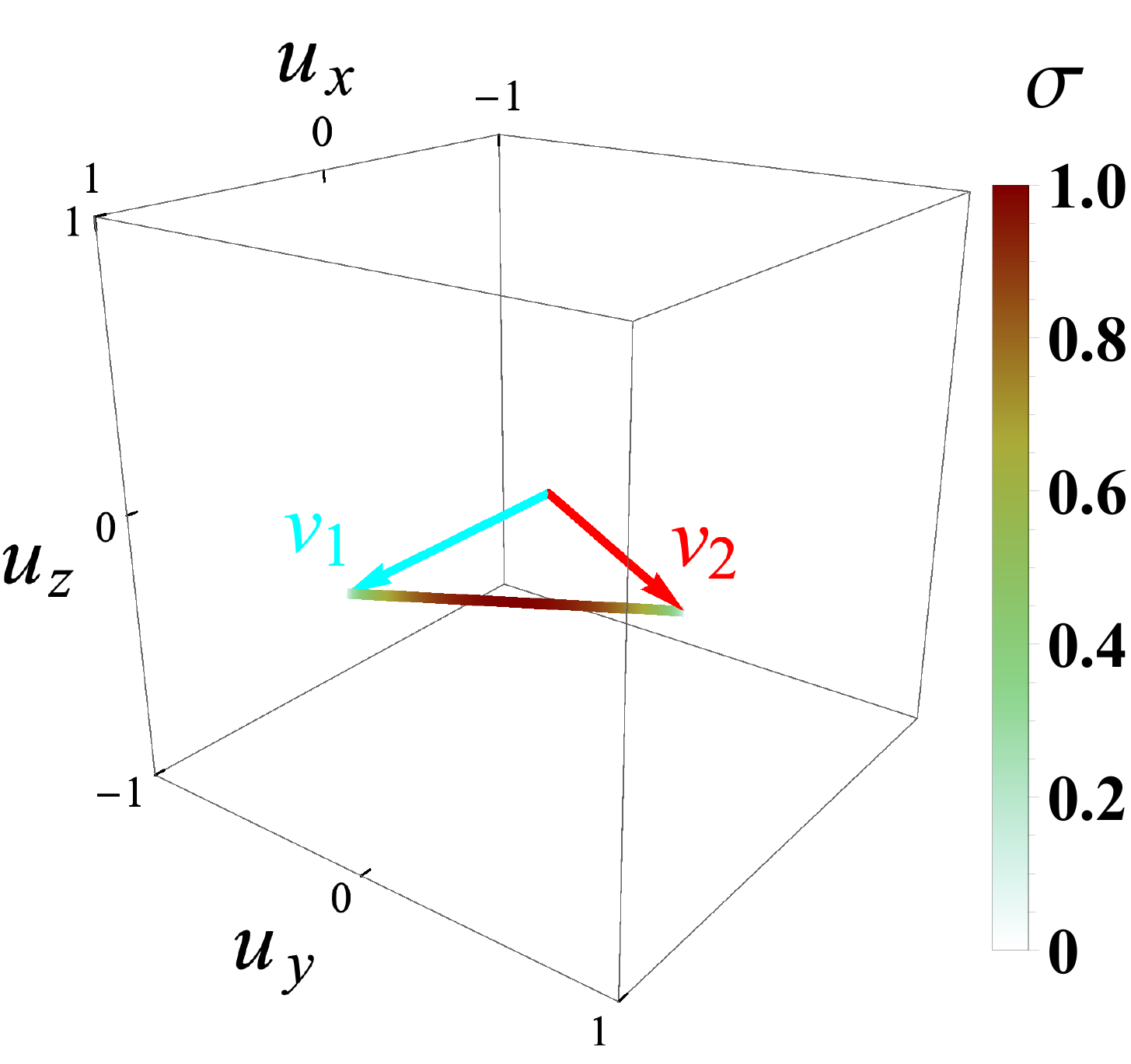}
    \caption{Growth rate $\sigma$ for the velocity of frame $\vec{u}$ when $\epsilon = -1$. We note that the thickness of the line that indicates perturbation growth is actually infinitesimally small.}
    \label{fig:GrowthRateMap}
\end{figure}

A schematic picture of the growth rate for each frame with velocity $\vec{u}$ is shown in Fig.~\ref{fig:GrowthRateMap}.
In the 4-dimensional case, a negative value of $\epsilon$, which corresponds to the case in which crossings exist in the ELN angular distribution, is necessary for instability, as in the 2-dimensional case.
The most important feature is that for $\epsilon<0$, the delta function in Eq.~(\ref{eq:G}) constrains the velocity of the frame in which the system is absolutely unstable on $\vec{u}-\vec{V}\parallel \vec{v}$.
Thus, an absolute instability appears only for $\vec{V}=\vec{0}$, and the instability is convective otherwise (see Table~\ref{tab:InstabilityCondition}).
These results suggest that the instabilities grow toward the direction between the two neutrino beams.
It may be interpreted intuitively as follows;
a perturbation generated at $(t, \vec{x}) = (0, \vec{0})$ can travel with the velocity $\vec{v}_1$ and $\vec{v}_2$ at each point;
therefore, at time $t$, all the points the perturbation can reach are $\vec{x} = (s\vec{v}_1 + (1 - s)\vec{v}_2)t$ with $s \in [0, 1]$, which is consistent with the support of Eq.~(\ref{eq:G}).
\begin{table}[htb]
    \caption{Associations between instabilities and model parameters for 4-dimensional perturbations.}
    \begin{ruledtabular}
        \begin{tabular}{cc}
            Classification & Condition\\
            \hline
            Stable & $\epsilon\geq 0$ \\
            Convectively unstable & $\epsilon<0,\ \vec{V}\neq  0$\\
            Absolutely unstable & $\epsilon<0,\ \vec{V}=0$ \\
        \end{tabular}
    \end{ruledtabular}
    \label{tab:InstabilityCondition}
\end{table}

For CCSNe, it has been pointed out that ELN crossings may occur in decoupling regions \cite{Capozzi2018a}.
Usually, $\nu_e$ overwhelms $\bar{\nu}_e$ in almost all directions, and $\bar{\nu}_e$ dominates only for almost radial directions in this region.
This situation may be approximated by the 2-beam model with a radial antineutrino beam with $\mathscr{G}_1<0$ and $\vec{v}_1 = \vec{e}_r$ and a neutrino beam in another direction with $\mathscr{G}_2>0$ and $\vec{v}_2\neq~\vec{e}_r$, where $\vec{e}_r$ is the unit vector parallel to the radial direction.
Therefore, flavor conversion occurs at least toward the radial direction, which may affect the heating of the stalled shock.

In addition, the possibility of ELN crossings in the preshock region was suggested in our previous study~\cite{Morinaga2020a}.
Because the average energy of $\bar\nu_e$ is larger than that of $\nu_e$, $\bar\nu_e$ scattered on nuclei in the preshock region can overwhelms $\nu_e$ while $\nu_e$ usually does $\bar\nu_e$ in the radial direction.
We concluded that the flavor instability does not propagate into the postshock region from the group velocity of the unstable modes in 1 spatial dimension.
The results shown in this study, however, might overturn this conclusion.
Radially-peaked $\nu_e$ distribution seems to be imitated by a beam with $\mathscr{G}_1 > 0$ and $\vec{v_1} = \vec{e}_r$ in the 2-beam model while the scattered $\bar\nu_e$ distribution does by a radially-ingoing beam with $\mathscr{G}_2 < 0$ and $\vec{v_2} = -\vec{e}_r$.
As a result, from Eq.~(\ref{eq:G}), perturbations can grow both outward and inward, which leads flavor conversions in the postshock region. 
In other words, scattered neutrinos seems to convey the instability into the postshock region as the intuitive interpretation mentioned above.
Realistically, however, neutrinos do not have a discrete spectrum, as in the 2-beam model, but rather a continuous one.
Further investigations should be conducted to determine how instabilities propagate in realistic supernovae.

\subsection{Continuous distribution}
Although we should consider a continuous spectrum in general, as mentioned above, it seems difficult to obtain the DR $\Delta(k)\equiv\det\mat{D}(k)=0$ because $\mat{D}$ has uncountably infinite dimensions, and it seems necessary to calculate the functional determinant of $\mat{D}$.
It should be noted that spectral discretization generates many spurious modes, which may affect the features of the instability, and thus, we cannot disregard the continuous nature of the spectrum \cite{Sarikas2012, Morinaga2018}.
Fortunately, we can prove that the DR is given by the determinant of a $4\times 4$ matrix called the polarization tensor and that the instability depends only on this determinant.

The Green's function for the linearized equation in Eq.~(\ref{eq:LinearizedKineticFast}) is defined as
\begin{align}
    &v\cdot \left\{i\partial -\Lambda_\mathrm{c}(x)\right\}G_{\vec{v}\vec{v}'}(x) \nonumber\\
    &+ \int \frac{d\vec{v}''}{4\pi} \mathscr{G}(\vec{v}'')v\cdot v''G_{\vec{v}''\vec{v}'}(x) = \delta^4(x)\delta(\vec{v},\vec{v}').
\end{align}
By defining $a_{\vec{v}}^\mu(k) \equiv \int \frac{d^2\vec{v}'}{4\pi}\mathscr{G}(\vec{v}')v'{}^\mu G_{\vec{v}'\vec{v}}(k+\Lambda_\mathrm{c})$, the Green's function $G$ can be expressed as
\begin{align}
    G_{\vec{v}\vec{v}'}(k+\Lambda_\mathrm{c}) = \dfrac{\delta(\vec{v},\vec{v}')-v\cdot a_{\vec{v}'}(k)}{v\cdot k}
    \label{eq:Gvv}
\end{align}
in a locally constant and uniform approximation of the potential $\Lambda_\mathrm{c}(x)$.
By substituting this expression into the definition of $a^\mu_{\vec{v}}$, we obtain the following linear equation for $a^\mu_{\vec{v}}$:
\begin{align}
    \Pi^{\mu\nu}(k)a_{\vec{v}\nu}(k) = \mathscr{G}(\vec{v})\frac{v^\mu}{v\cdot k},
\end{align}
where
\begin{align}
     \Pi^{\mu\nu}(k) \equiv \eta^{\mu\nu} + \int \frac{d\vec{v}}{4\pi} \mathscr{G}(\vec{v})\dfrac{v^\mu v^\nu}{v\cdot k}
\end{align}
is called the polarization tensor~\cite{Izaguirre2017}.
By substituting this into Eq.~(\ref{eq:Gvv}), the following expression for the Green's function is derived:
\begin{align}
    G_{\vec{v}\vec{v}'}(k+\Lambda_\mathrm{c}) = \dfrac{\delta(\vec{v},\vec{v}')}{v\cdot k} - \dfrac{v^\mu \Pi^{-1}_{\mu\nu}(k) v'{}^\nu}{(v\cdot k)(v'\cdot k)}\mathscr{G}(\vec{v}').
\end{align}

Therefore, all of the singularities of $G$ originate from the roots of $\det\Pi(k-\Lambda_\mathrm{c}) = 0$ and $v\cdot (k-\Lambda_\mathrm{c}) = 0$ for arbitrary $\vec{v}$.
The latter does not influence the instability because $\omega$ is real for all $\vec{k}\in\mathbb{R}^3$.
Thus, it is sufficient to investigate the instability for the DR $\det\Pi(k)=0$.

\section{Conclusion}
In this paper, we have conducted a spatiotemporal linear instability analysis of collective neutrino flavor conversion in 4-dimensional spacetime.
We have analytically derived the asymptotic form of the linear perturbations for a 2-beam model and
highlighted the importance of 4-dimensional perturbations for collective neutrino oscillations.
In addition, we have shown that collective neutrino oscillations might influence shock heating not only in the decoupling region but also in the preshock region of a CCSN.
We have also discussed the application of the method of instability analysis to continuous spectra, which are more realistic in nature.

As many previous studies have pointed out, ELN crossings allow neutrino flavor conversion, and the spatiotemporal behavior of such conversion in 4-dimensional spacetime is revealed in this paper for the first time.
When we consider 2-dimensional perturbations, the instabilities can appear as both absolute and convective depending on the directions chosen for the perturbations, and such an analysis cannot elucidate how perturbations truly behave in 4-dimensional spacetime.
When 4-dimensional perturbations are taken into account, it is clarified that the instabilities are actually convective and the perturbations grow toward the direction between the neutrino and antineutrino beams.

Needless to say, one of our goals is to gain an understanding of nonlinear spatiotemporal behaviors.
However, it seems that more qualitative studies are also needed for guidance, especially in the current situation, in which numerical calculations seem to be extremely difficult.
Additionally, more systematic studies are certainly needed to determine how flavor instabilities are actually triggered and evolve in supernovae.
It is true that the 2-beam model is simply a toy model, but we hope that this first-ever study on the spatiotemporal behaviors of flavor conversion in 4-dimensional spacetime will help develop an understanding of the more realistic flavor conversion behaviors in supernovae.

\begin{acknowledgments}
I am grateful to Shoichi Yamada and Hiroki Nagakura for useful discussions.
I am supported by a JSPS Grant-in-Aid for JSPS Fellows (No. 19J21244) from the Ministry of Education, Culture, Sports, Science and Technology (MEXT), Japan.
\end{acknowledgments}


%

\end{document}